\documentclass[11pt,twoside]{article}

\usepackage{asp2014}
\usepackage{graphicx}
\usepackage{sidecap}
\aspSuppressVolSlug
\resetcounters

\begin{document}

\title{Xova: Baseline-Dependent Time and Channel Averaging for Radio Interferometry}

\author{Marcellin~Atemkeng$^{1}$, Simon~Perkins$^{2}$, Jonathan~Kenyon$^{1}$,  Benjamin~Hugo$^{2,1}$, and Oleg~Smirnov$^{1,2}$}
\affil{$^1$Rhodes University, Makhanda (Grahamstown), Eastern Cape, South Africa}
\affil{$^2$South African Radio Astronomy Observatory, Cape Town, Western Cape, South Africa}

\paperauthor{Marcellin~Atemkeng}{m.atemkeng@gmail.com}{}{Rhodes University}{Radio Astronomy Techniques and Technologies}{Makhanda (Grahamstown)}{Eastern Cape}{6140}{South Africa}
\paperauthor{Simon~Perkins}{simon.perkins@gmail.com}{}{South African Radio Astronomy Observatory}{Radio Astronomy Research Group}{Cape Town}{Western Cape}{7925}{South Africa}
\paperauthor{Jon~Kenyon}{}{}{Rhodes University}{Radio Astronomy Techniques and Technologies}{Makhanda (Grahamstown)}{Eastern Cape}{6140}{South Africa}
\paperauthor{Ben~Hugo}{}{}{South African Radio Astronomy Observatory}{Radio Astronomy Research Group}{Cape Town}{Western Cape}{7925}{South Africa}
\paperauthor{Oleg~Smirnov}{osmirnov@gmail.com}{}{Rhodes University}{Radio Astronomy Techniques and Technologies}{Makhanda (Grahamstown)}{Eastern Cape}{6140}{South Africa}



\begin{abstract}
Xova is a software package that implements baseline-dependent time and channel averaging on Measurement Set data. The $uv$-samples along a baseline track are aggregated into a bin until a specified decorrelation tolerance is exceeded. The degree of decorrelation in the bin correspondingly determines the amount of channel and timeslot averaging that is suitable for samples in the bin. This necessarily implies that the number of channels and timeslots varies per bin and the output data loses the rectilinear input shape of the input data.
\end{abstract}

\section{Effects of Time and Channel  Averaging}
Consider  $\mathcal{V}_{pq}=\mathcal{V}(\mathbf{u}_{pq}(t,\nu))$ as the visibility sampled by the baseline $pq$ at time $t$ and frequency $\nu$. An interferometer is non-ideal in the
sense that the measured visibility is the average of this sampled visibility over the sampling bin, $B_{kr}^{[\Delta t \Delta \nu]}=[t_k-\Delta t/2, t_k+\Delta t/2]\times [\nu_r-\Delta \nu /2, \nu_r+\Delta \nu/2]$:
\begin{equation}
\widetilde{\mathcal{V}}_{pq}=\frac{1}{\Delta t \Delta \nu} \iint\limits_{B_{kr}^{[\Delta t \Delta \nu]}} \mathcal{V}^{}(\mathbf{u}_{pq}(t,\nu)) \text{d}t \text{d}\nu,\label{eq0}
\end{equation}
where $\Delta t$ and $\Delta \nu$ are the integration intervals. If $\Pi_{pq}$ represents a normalized 2D top-hat window for a baseline $pq$ then Eq. (1) is equivalent to the infinitesimal integral:
\begin{equation}
\widetilde{\mathcal{V}}_{pq}= \iint\limits_{\infty}  \Pi_{pq}(t-t_k, \nu-\nu_r)\mathcal{V}_{pq}(t,\nu) \text{d}t \text{d}\nu,\label{eq1}
\end{equation}
which is a convolution in the Fourier space, i.e.:
\begin{alignat}{2}
\widetilde{\mathcal{V}}_{pq}&=[ \Pi_{pq}\circ\mathcal{V}_{pq}](\mathbf{u}_{pq}(t_k,\nu_r))\\
&= \delta _{pqkr}[\Pi_{pq}\circ\mathcal{V}_{pq}],\label{eq2}
\end{alignat}
where $\delta_{pqkr}(\mathbf{u})=\delta (\mathbf{u}-\mathbf{u}_{pqkr})$ is the Delta function shifted to the sampled point $pqkr$. For an observation with frequency range $F=\Delta \nu \times N_{\nu}$ and total observing period $T=\Delta t \times N_{t}$, observing for long times and large frequency
ranges leads to storage issues, as well as computation cost since $\{ T, F\} \propto \{N_t, N_{\nu}\}$ if $\Delta t$ and $\Delta \nu$ most remain sufficiently small. For aggressive averaging, $\Delta t$ and $\Delta \nu$ are large which makes $\Pi_{pq}$ to deviate significantly from $\delta_{pqkr}$ and  therefore causes the visibility to decorrelate:
$\mathcal{V}_{pq}\neq \widetilde{\mathcal{V}}_{pq}$.
To derive the effect of averaging on the image, we can reformulate
Eq.~\ref{eq2} as:
\begin{alignat}{2}
\widetilde{\mathcal{V}}_{pq}&=\mathcal{F}\{\mathcal{P}_{pqkr}\}\big(\Pi_{pq}\circ\mathcal{F}\{\mathcal{I}\} \big),\label{eq3}
\end{alignat}
where the apparent sky $\mathcal{I}$ is the inverse Fourier transform of $\mathcal{V}_{pq}$ and $\mathcal{P}_{pqkr}$ is the inverse Fourier transform of $\delta_{pqkr}$. Here $\mathcal{F}$  represents the Fourier transform. Inverting the sum of Eq.~\ref{eq3} over all the baselines results in
an estimate of the sky image:
\begin{alignat}{2}
\widetilde{\mathcal{I}}&=\sum_{pqkr}W_{pqkr}\mathcal{P}_{pqkr}\circ \big(\mathcal{D}_{pqkr}\mathcal{I}\big),\label{eq4}
\end{alignat}
where $W_{pqkr}$ is the weight at the sampled point $pqkr$. We note that the apparent sky $\mathcal{I}$ is now tapered by the baseline-dependent distortion distribution $\mathcal{D}_{pqkr}$, the latter being the inverse Fourier transform of the baseline-dependent top-hat window:
\begin{alignat}{2}
\mathcal{D}_{pqkr} &= \mathcal{F}^{-1}\{\Pi_{pq}\}
                    &=\mathrm{sinc}\left(\frac{\Delta \Psi}{2}\right) \mathrm{sinc}\left(\frac{\Delta \Phi}{2}\right).
\end{alignat}
For a source at the sky location $\mathbf{l}$, the $\Delta \Psi$ and $\Delta \Phi$ are the phase difference  in time and frequency, respectively:
\begin{alignat}{2}
\Delta \Psi&=2\pi\Delta \mathbf{u}_{pq}(t, \nu_r)\mathbf{l};
\Delta \Phi &= 2\pi\Delta \mathbf{u}_{pq}(t_k, \nu)\mathbf{l}.
\end{alignat}
Assuming no other corruption effects apart from decorrelation and assuming naturally weighting a sky with a single source; with decorrelation in effect Eq. \ref{eq4} becomes:
\begin{alignat}{2}
\widetilde{\mathcal{I}}&=\mathcal{P}_{pqkr}\circ \big(\mathcal{D}_{pqkr}\mathcal{I}\big)
&=\mathcal{D}_{pqkr}\mathcal{I}.\label{eq4x}
\end{alignat}
 Note that in this formulation, we have assumed that $\mathcal{P}_{pqkr}=\delta(\mathbf{l})$ is baseline independent as opposed to \citet{atemkeng2020fast}. Eq. \ref{eq4x} is simulated in Figure 1 using the MeerKAT telescope at 1.4 GHz showing the apparent intensity of a 1 Jy source, as seen by the shortest baseline, medium-length baseline, and longest baseline, as a function of distance from the phase center.   We see that decorrelation is severe on the longest baseline then followed by the medium length baseline, and that decorrelation is a function of source position in the sky.
\begin{SCfigure}
  \centering
  \caption{ Effect of time averaging: The data is sampled at 1 s and 84 kHz frequency resolutions then averaged only in time across 15 s. }
  \includegraphics[width=0.6\textwidth]%
    {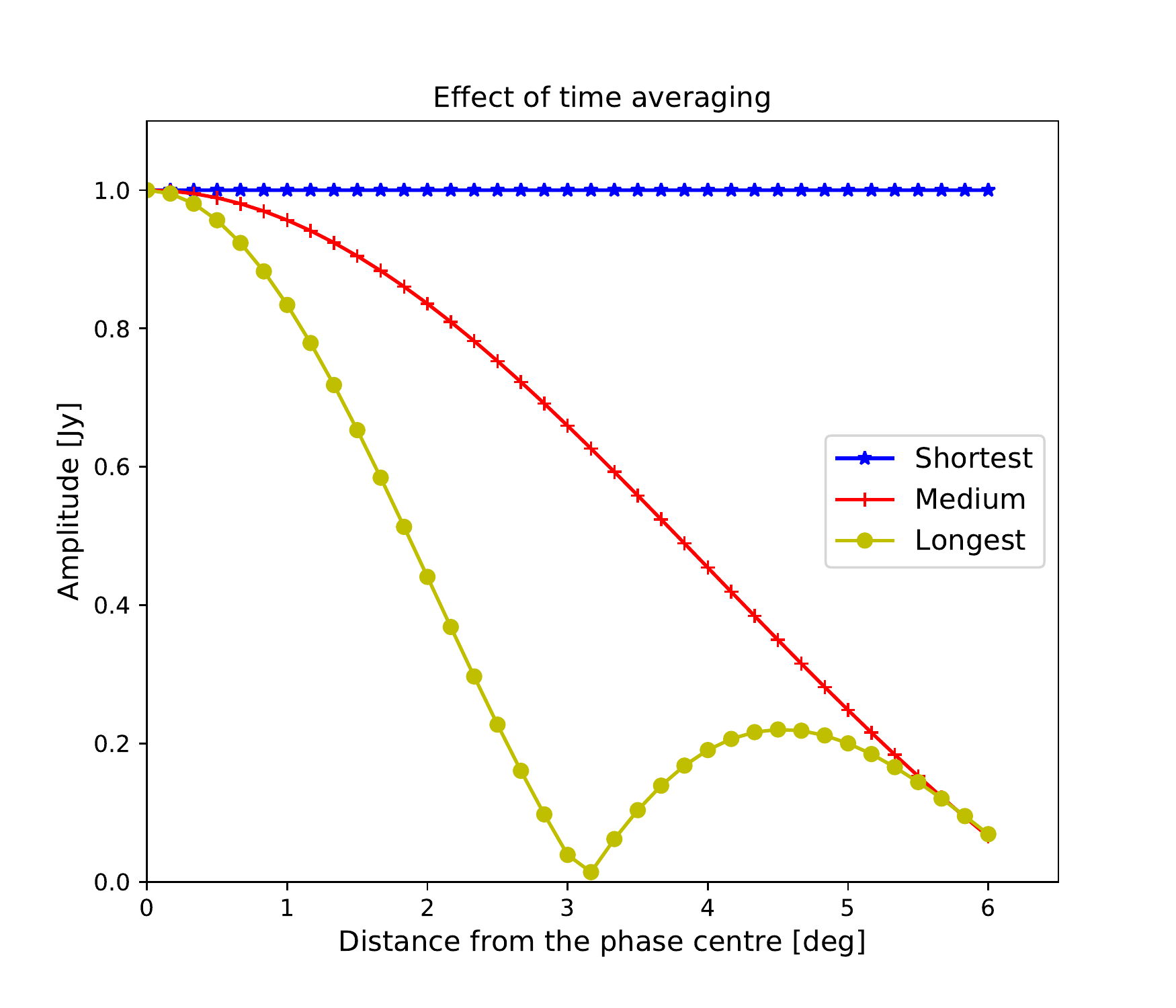}
\end{SCfigure} 
 \section{Baseline-Dependent Time and Channel  Averaging (BDA)}
 The distortion distribution, $\mathcal{D}_{pqkr}$ depends on each baseline and its rotation orientation in the Fourier space which makes the decorrelation to be baseline-dependent. For decorrelation to be baseline-independent, the rectangular  sampling bin $B_{kr}^{[\Delta t \Delta \nu]}$  across which the data is averaged must be kept baseline-dependent as opposed to the fixed  sampling bin currently employed in radio interferometer correlators:
\begin{alignat}{2}
B_{kr}^{[\Delta_{\mathbf{u}_{pq}} t \Delta_{\mathbf{u}_{pq}} \nu]}=[t_k-\Delta_{\mathbf{u}_{pq}} t/2, t_k+\Delta_{\mathbf{u}_{pq}} t/2]\times [\nu_r-\Delta_{\mathbf{u}_{pq}} \nu /2, \nu_r+\Delta_{\mathbf{u}_{pq}} \nu/2],
\end{alignat}
where the integration intervals $\Delta_{\mathbf{u}_{pq}} t$ and $\Delta_{\mathbf{u}_{pq}} \nu$ are now also baseline-dependant. In this case Eq.~\ref{eq4}  becomes:
\begin{alignat}{2}
\widetilde{\mathcal{I}}&=\sum_{pqkr}\mathcal{W}_{pqkr}\mathcal{P}_{pqkr}\circ \big(\mathcal{D}_{}\mathcal{I}\big),\label{eq5}
\end{alignat}
where $\mathcal{D}_{}=\mathcal{D}_{pqkr}=\mathcal{D}_{\alpha \beta kr}$ is the distortion distribution, which is now equal across all the baselines, $pq$ and $\alpha \beta$ no matter their orientation. We provide details on the implementation of $\mathcal{D}_{}$ in Sections~\ref{section3} and \ref{section4}.

\section{Technologies}
\label{section3}
The core of Xova's BDA algorithm is implemented using two recent parallelisation and acceleration frameworks: (1) \textbf{Dask} \citep{dask-2015} is a Python parallel computing library that expresses programs as \textbf{Computational Graphs} whose individual tasks are scheduled on multiple cores or nodes.  \textbf{Dask collections} abstract underlying graphs with familiar \textbf{Array} and \textbf{Dataframe} interfaces. (2) \textbf{Numba} \citep{numba-2015} a JIT compiler that translates the BDA algorithm, expressed as a subset of Python and NumPy code, to accelerated machine code. These are implemented in Xova as follows: \textbf{dask-ms} \citep{O8-131_adassxxx} exposes Measurement Set columns as dask arrays for ingest by Xova then \textbf{Codex Africanus} \citep{O8-131_adassxxx} a Radio Astronomy Algorithms Library, applies BDA, implemented in numba to dask arrays, producing averaged dask arrays and  \textbf{dask-ms} writes the averaged dask arrays to a new Measurement Set.

\section{Xova}
\label{section4}
\begin{SCfigure}
  \centering
  \caption{ The parts of the baseline closer to the phase centre are subject to greater averaging }
  \includegraphics[width=0.6\textwidth]%
    {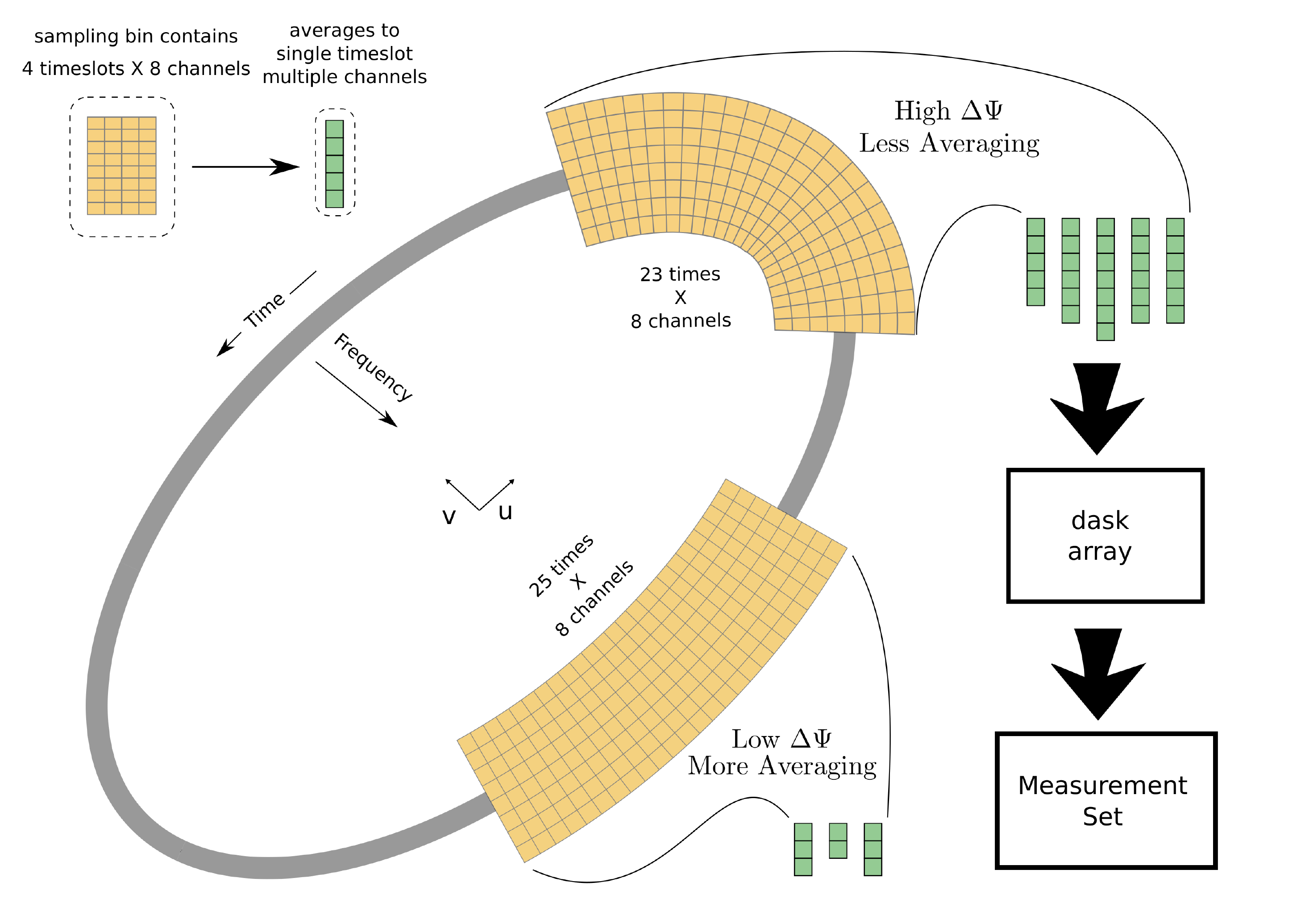}
\end{SCfigure}
For each baseline (See Figure 2): Measurement Set timeslots are aggregated into averaging bins until $\textrm{sinc}\left(\Delta \Psi / 2\right)$ falls below decorrelation tolerance $\mathcal{D}$. The acceptable corresponding change in frequency $\Delta \Phi = 2 \, \textrm{sinc}^{-1}\left(\mathcal{D} / \textrm{sinc}\left(\Delta \Psi / 2\right)\right)$ is calculated and channel width $\Delta \nu$ is derived from $\Delta \Phi$ and used to divide the original band into a new channelisation.

\section{Results}
Figure~3 shows the image of a high-resolution data set imaged without BDA (right panel) and with $95\%$ decorrelation tolerance BDA (left panel).  We note that BDA does not distort the image when compared to the no BDA image. 
\articlefigure[width=.9\textwidth]{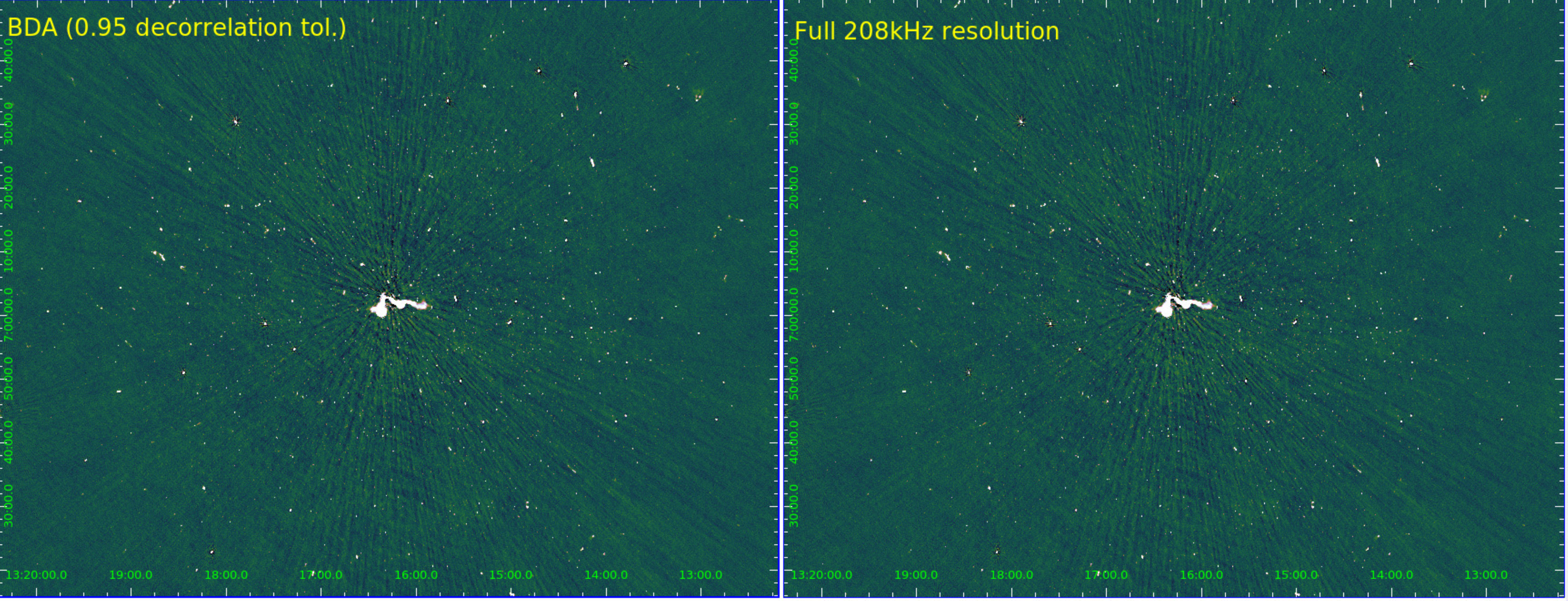}{bda-images}{BDA (left) vs. no BDA (right).}

\acknowledgements   The research of Oleg Smirnov is supported by the South African Research Chairs Initiative of the Department of Science and Technology and National Research Foundation.

\bibliography{P2-246}

\begin{thebibliography}{}
\expandafter\ifx\csname natexlab\endcsname\relax\def\natexlab#1{#1}\fi
\expandafter\ifx\csname url\endcsname\relax
  \def\url#1{\texttt{#1}}\fi
\expandafter\ifx\csname urlprefix\endcsname\relax\def\urlprefix{URL }\fi
\providecommand{\eprint}[2][]{\url{#2}}

\bibitem[{Atemkeng et~al.(2020)Atemkeng, Smirnov, Tasse, Foster, \&
  Makhathini}]{atemkeng2020fast}
Atemkeng, M., Smirnov, O., Tasse, C., Foster, G., \& Makhathini, S. 2020,
  Monthly Notices of the Royal Astronomical Society, 499, 292

\bibitem[{Lam et~al.(2015)Lam, Pitrou, \& Seibert}]{numba-2015}
Lam, S.~K., Pitrou, A., \& Seibert, S. 2015, in Proceedings of the Second
  Workshop on the LLVM Compiler Infrastructure in HPC (New York, NY, USA:
  Association for Computing Machinery), LLVM '15.
  \urlprefix\url{https://doi.org/10.1145/2833157.2833162}

\bibitem[{{Perkins} et~al.(2021)}]{O8-131_adassxxx}
{Perkins}, S.~J., et~al. 2021, in ADASS XXX, edited by J.-E. {Ruiz}, \&
  F.~{Pierfederici} (San Francisco: ASP), vol. TBD of ASP Conf. Ser., 999 TBD

\bibitem[{Rocklin(2015)}]{dask-2015}
Rocklin, M. 2015, in Proceedings of the 14th Python in Science Conference,
  edited by K.~Huff, \& J.~Bergstra, 130

\end{thebibliography}


\end{document}